\documentclass[journal]{IEEEtran}

\usepackage{multicol}
\usepackage{multirow}
\usepackage{graphicx}
\usepackage{subcaption}
\usepackage{amsmath}
\usepackage{amssymb}

\usepackage{subfiles}
\usepackage{xcolor,soul}
\definecolor{mypink1}{rgb}{0.858, 0.188, 0.478}

\definecolor{myhlcolor}{gray}{0.85}
\sethlcolor{yellow}
\definecolor{myOrange}{rgb}{1,0.5,0}
\definecolor{myGreen}{rgb}{.2,.9,.2}

\usepackage{scalerel}
\usepackage{tikz}
\usetikzlibrary{svg.path}
\definecolor{orcidlogocol}{HTML}{A6CE39}
\usepackage{verbatim}
\usepackage{scalerel}
\usepackage{tikz}
\usetikzlibrary{svg.path}

\definecolor{orcidlogocol}{HTML}{A6CE39}
\tikzset{
	orcidlogo/.pic={
		\fill[orcidlogocol] svg{M256,128c0,70.7-57.3,128-128,128C57.3,256,0,198.7,0,128C0,57.3,57.3,0,128,0C198.7,0,256,57.3,256,128z};
		\fill[white] svg{M86.3,186.2H70.9V79.1h15.4v48.4V186.2z}
		svg{M108.9,79.1h41.6c39.6,0,57,28.3,57,53.6c0,27.5-21.5,53.6-56.8,53.6h-41.8V79.1z M124.3,172.4h24.5c34.9,0,42.9-26.5,42.9-39.7c0-21.5-13.7-39.7-43.7-39.7h-23.7V172.4z}
		svg{M88.7,56.8c0,5.5-4.5,10.1-10.1,10.1c-5.6,0-10.1-4.6-10.1-10.1c0-5.6,4.5-10.1,10.1-10.1C84.2,46.7,88.7,51.3,88.7,56.8z};
	}
}

\newcommand\orcidicon[1]{\href{https://orcid.org/#1}{\mbox{\scalerel*{
				\begin{tikzpicture}[yscale=-1,transform shape]
				\pic{orcidlogo};
				\end{tikzpicture}
			}{|}}}}

\usepackage[bookmarks=false]{hyperref} 
\PassOptionsToPackage{bookmarks=false}{hyperref}

\ifCLASSINFOpdf
\else
\fi

\begin{document}

\title{An Experimental Validation of Precise GNSS Time Synchronization in Vehicular Networks}

\author{Khondokar Fida Hasan\textsuperscript{\orcidicon{0000-0002-8008-8203}},~\IEEEmembership{Member,~IEEE,}~Yanming Feng\textsuperscript{\orcidicon{0000-0001-6548-3347}},~\IEEEmembership{}and~Yu-Chu Tian\textsuperscript{ \orcidicon{0000-0002-8709-5625}},~\IEEEmembership{Senior Member,~IEEE}

\thanks{Manuscript received ???? ; revised ????. This work was supported in part by the Australian Research Council (ARC) through the Discovery Project Scheme under Grant DP170103305.} 
                
\thanks{The authors are with the School of Computer Science, Queensland University of Technology, GPO Box 2434, Brisbane QLD 4001, Australia.} 

\thanks{Corresponding author: Khondokar Fida Hasan (fida.hasan@qut.edu.au)}

\thanks{Color versions of one or more of the figures in this paper are available
online at http://ieeexplore.ieee.org.}
\thanks{Digital Object Identifier ???}
}

\markboth{Submitted to IEEE Transaction for possible publication. Copyright may be transferred without notice 
	 \hfill }{}

\maketitle

\begin{abstract}
Time Synchronization utilizing the Global Navigation Satellite System (GNSS) is being increasingly investigated for vehicular networks. The availability and accuracy of GNSS timing solutions in varying road environments are a recognised challenge due to GNSS signal blockages. Considering vehicular networks, this paper conducts a systematic analysis of the required time synchronization, and develops a GNSS-based time synchronization solution. It demonstrates the availability and capability of GNSS time synchronization experimentally using commercial-grade GNSS receivers and off-the-shelf communication devices. Our experiments show that the timing accuracy can be as good as 2 microseconds. A momentary complete outage of GNSS time solution due to signal blockage on the road can lead to a clock error of up to sub-10 microseconds, which still meets the desired requirement for most vehicular applications.
\end{abstract}


\IEEEpeerreviewmaketitle

\section{Introduction}

\IEEEPARstart{V}{ehicular} network is one of the enabling technologies for Cooperative Intelligent Transportation Systems (C-ITS). It provides wide-scale use cases, including time-sensitive ones. For example, some road safety messages that enable cooperative sensing and maneuvers such as Lane Change Assistance, Intersection Control, and Collision Avoidance, requiring an end-to-end transmission latency of less than 100 ms ~\cite{Boban2018,karagiannis2011vehicular}. Accurate timing of these transmitted safety messages is, therefore, critical to meet this requirement. Time synchronization is a proven mechanism to attain accurate time across a vehicular network~\cite{FidaHasan2018}. 

Among a variety of Vehicle-to-Everything (V2X)- support access technologies, Dedicated Short Range Communication (DSRC), and Long-Term Evolution (LTE) are two promising candidates~\cite{MacHardy2018,Seo2016}. 
Due to the highly dynamic nature of vehicular nodes ($\approx$ 200 Km/h relative speed), all nodes under a short-range Vehicular Ad-hoc Network (VANET) stay each others' communication range only for a fraction of a second. As a result, the topology of the network changes rapidly, making DSRC-enabled message-transfer-based time synchronization 
ineffective \cite{Xu2004,Mahmood2016a,Mahmood2016}.
Long Term Evolution (LTE), on the other hand, covers a wide area of service connectivity. However, it is still unclear what the timing performance is for the end nodes interconnected to a cellular network because reports on this topic have not been found. 
This motivates us to use GNSS service for time synchronization in vehicular networks as GNSS is a recognized source of accurate and precise timing information and has been used in many distributed systems.  Moreover, GNSS-based external time transfer does not consume data communication bandwidth (e.g. DSRC or LTE Channels) for transferring synchronization messages among nodes, thus creating zero overhead to the data communication bandwidth~\cite{Defraigne2013,gps2006book}. 

Despite the zero-overhead and nano-second level timing accuracy offered by GNSS services, the achievable clock accuracy to an end device can vary widely depending on the factors such as the type of GNSS receivers, the hardware's quality with interfacing-technique of the application device, and the availability of GNSS signals~\cite{Piester2008}.  For example, standalone consumer-grade GNSS receivers that are prevalent in modern vehicles for route navigational support uses low-end inexpensive GNSS modules and antennas, which may not offer the same performance of high-end expensive receivers operating in a precise positioning and timing mode. In addition, one can assume that vehicular networks are exclusively outdoor-based networks where Line of Sight (LOS) propagation is uninterrupted and GNSS signals are always available. In reality, there are, however, obstacles due to trees and urban road scenarios such as high-rising buildings and tunnels limiting the signal reception. Therefore, the challenge is to achieve the required synchronization accuracy in real-world road environments via low-end receivers that are popularly used in modern vehicles. In this research, we leverage the advancement of multi-GNSS receivers that can receive satellite signals from all active constellations, thus increasing the signal coverage or solution availability. The research is significant in realizing GNSS capacity to provide the required time synchronization for vehicular networks.

This paper studies the attainability and capacity of the GNSS timing service to vehicular networks by utilizing low-cost consumer-grade GNSS receivers and off-the-shelf communication devices. 
It conducts a systematic analysis of the required time synchronization for vehicular networks. Then, it presents a GNSS-based precise time synchronization solution for vehicular networks. While developing the solution the timing capacity of GNSS is determined in three stages: (i) the operational stability and accuracy of GNSS steering time from NMEA (National Marine Electronics Association) sentence data, (ii) the accuracy of the receiver generated electrical signal pulse named Pulse Per Second (PPS) signal, that repeats precisely once per second, and (iii) the stability and accuracy of time by combining NMEA and PPS. Overall, conducting error analysis on a series of field tests, this paper demonstrates that using GNSS service, the attainable timing accuracy is better than existing DSRC- and LTE-based timing services.

The paper is organized as follows: Section II defines the requirements for time synchronization in VANET. Section III analyzes the synchronization problem with DSRC-based communication technologies for vehicular networks. GNSS time synchronization based development and its performance evaluation are presented in Section IV. Then, a quick experiment on LTE-based network timing accuracy is presented in Section V. Finally, Section VII concludes the paper.


\section{Requirements for Synchronization Accuracy}

Several time synchronization mechanisms 
have been proposed and implemented over the years in computer and communication networks~\cite{johannessen2004time,mills2016computer,bregni2002synchronization,wu2010clock}. However, most often they are ill-suited and/or incompatible with vehicular network requirements \cite{8930528}.
Time synchronization for vehicular networks 
needs to address three criteria: the degree of accuracy needed, i.e, the accuracy with respect to some external time reference; the degree of precision needed, i.e, the range of accuracy that can be maintained at the clock of a node; and the availability and longevity of the synchronization mechanism, i.e, whether the system needs to stay synchronized indefinitely or to establish synchronization on-demand. 

\subsection[Accuracy]{Accuracy Requirements} 
Timing accuracy refers to the closeness of the time of a node to a standard physical time, e.g., the universal coordinated time (UTC) or a standard local time. It is a measurement value resulting from the external synchronization of the node. 
Let \textit{V(t) $\subseteq N$} denote a set of vehicular nodes aligned with a physical time \textit{t}. 
Let \textit{p} and \textit{q} be a pair of nodes such that ${p,q \in N}$. 
Also, let $\alpha$ denote the required timing accuracy. Then, for each node the system should satisfy:
\begin{equation} 
\forall t \forall p \in V(t) : |C_p(t)-t| \leq \alpha 
\label{eq:accuracy}
\end{equation}
%
%
where $C_p(t)$ is the time maintained by the clock of node \textit{p}. In vehicular networks, most of the communication interactions involve time-based decisions. For example, the real-time status messages of vehicular networks carry data about the position and velocity, which need to be determined accurately and to be delivered precisely on time.

In addition, collisions on roads can be handled by maintaining relative positioning accuracy is about 10 cm \cite{Caporaletti2012}, which strictly depends on the timing accuracy of 3 ms \cite{Hasan2018}.  Communication with the outside world via the Internet or other means requires maintaining accurate physical time for meaningful and successful interaction.  Therefore, maintaining timing accuracy $\alpha $ with respect to a standard reference time is a prerequisite in vehicular networks.

\subsection[Precision]{Precision Requirement} 
The term ‘time precision’ pertains to the agreement and closeness of a set of results. In a network, it refers to the closeness of the times kept at two or more nodes. It is often called \textit{instantaneous precision} and is measured as the degree or boundary of difference between clocks. The precision requirement concerns the internal synchronization of nodes. Let $\beta$ denote the required time precision. Then, a precise system should satisfy:
\begin{equation} 
\forall t \forall p, q \in V(t) : |C_p(t)-C_q(t)| \leq \beta
\label{eq:precision}
\end{equation}
where $C_p(t)$ and $C_q(t)$ are the time of the node $p$ and $q$, respectively. 
Precision is the expression of a boundary of accuracy which is essentially a function of the Mean ($\bar{a}$) and Standard Deviation ($\sigma$) of a set of times (t).
\begin{equation}
\begin{cases}
\bar{a} = \frac{1}{N}\sum_{i=0}^{N-1} x_{i} \\ 
\sigma = \sqrt {\frac{1}{N-1}\sum_{i=0}^{N-1}{(x_{i}-\bar{a})^2}}  
\end{cases}
\label{eqn:3}
\end{equation}
where $x_i$ is the instantaneous time difference between nodes. For a single node synchronized with an external standard clock, Equation (\ref{eq:precision})  reverts to the accuracy equation (\ref{eq:accuracy}). This implies that a precise system should satisfy the specified accuracy requirement. 

As the spectrum for vehicular networks is limited (75 MHz), maintaining an accurate time will improve spectrum utilization. To avoid interference and accommodate time inaccuracies between nodes, a Guard Interval (GI), also known as a Guard Band, is used in 
communication channel coordination protocols. A longer GI is undesirable because it generates idle time in the transmission process, thus reducing data throughput. Research shows that in 802.11n a reduction of guard interval from 800ns to 400ns increases the data rate by 11\% \cite{4557042}. Thus, maintaining good timing precision reduces the necessity of an extended Guard Band.

\subsection[Availability]{Availability Requirement} 

Availability  
is generally defined as the ability to deliver services upon demand. It captures the continuity, quality, and functionality of the services. Along with accuracy and precision, the availability of time synchronization services is also important in vehicular networks. Since communication in a vehicular network is a progressive process, the underlying time synchronization processes need to be updated continuously.

\begin{figure}[tb!]
	\begin{center}
		\includegraphics[width=1.0\columnwidth,trim=0 5 0 5,clip]{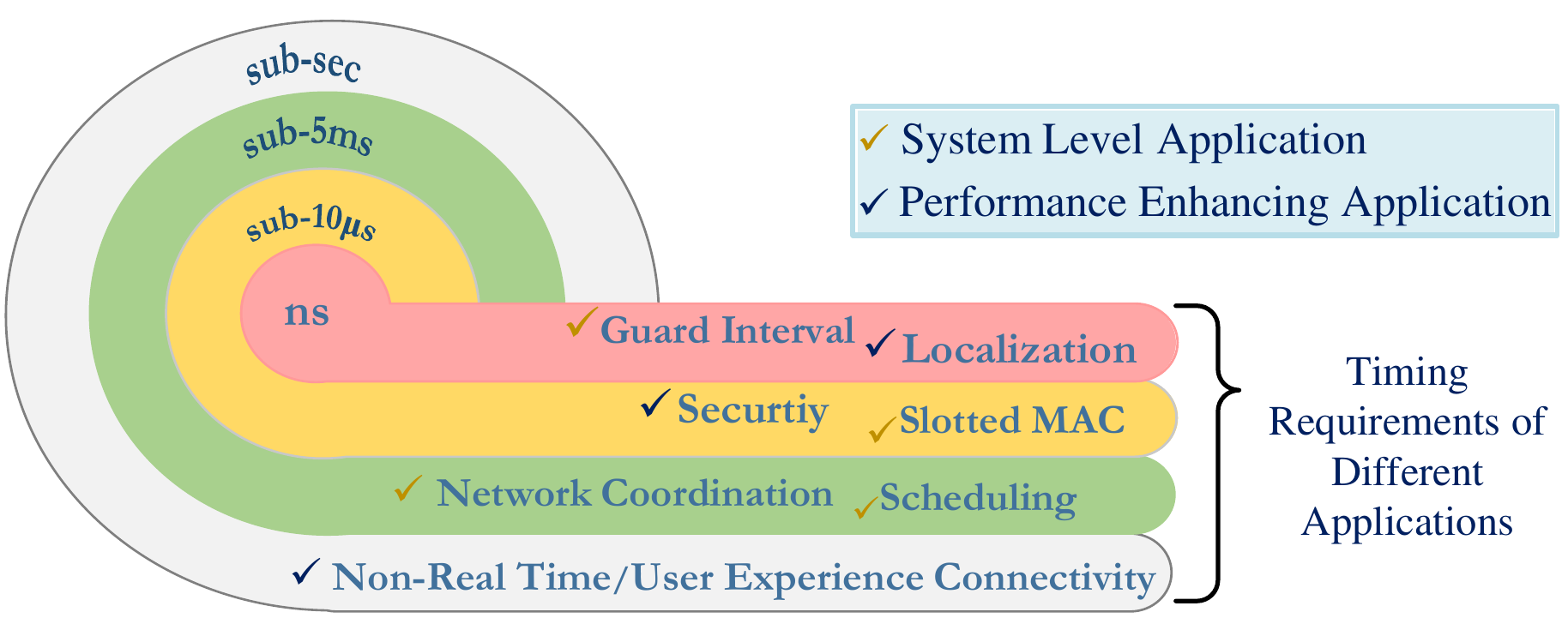}
	\end{center}
	\caption{A conceptual tier of requirements for time synchronization accuracy in vehicular network applications.}
	\label{fig:concept}
\end{figure}

Overall, from the requirements analysis of time synchronization conducted in \cite{Hasan2018} and \cite{FidaHasan2018}, \figurename~\ref{fig:concept} shows a conceptual tier of time synchronization requirements for different applications with respect to timing accuracy demand. The applications are divided into two categories: system level applications and Performance enhancing applications. They are presented within a range between nano-seccond to sub-seccond.

\section{Analysis of the Synchronization Problems with Current Proposed Technologies}

Synchronization in communication networks is a fundamental issue for data and network management  \cite{ghosh2011imple}. 
Over the years, the issue of time synchronization has been extensively investigated in the context of computer and telecommunication networks. Many protocols have been proposed and implemented to perform synchronization over computer and telecommunication networks. These protocols vary by the nature of the network, services, and precision of the timing requirements of different applications over the network. In vehicular networking, however, DSRC is a candidate technology capable of both Vehicle-to-Vehicle (V2V) and Vehicle-to-Infrastructure (V2I) communication, whereas LTE is more suitable for the applications of V2I communication \cite{Araniti2013}. This section presents a survey of different time synchronization protocols and the achievable accuracy of DSRC.

Existing DSRC-based time synchronization in VANET is defined is IEEE 802.11p, which is an amendment to the IEEE 802.11 WLAN standard. The 802.11 standard defines a Timing Synchronization Function (TSF) in MAC layer. TSF is a timer of 64-bit hardware counter with a resolution of 1 $\mu$s capable of performing 264 modulus counting. The timer runs by a local clock oscillator build on a WLAN chipset with a frequency accuracy of $\pm$0.01\%. 
TSF-based time synchronization relies on beacon messages from each node by using in-band frequency channels. TSF is feasible to synchronize the nodes in infrastructure mode.
In infrastructure mode, an Access Point (AP) broadcasts its TSF timer content into the network. Upon the reception, the nodes update their TSF from the time value of the received message. Since there are no access points in ad-hoc mode, time synchronization is developed in a distributed fashion where each node transmits beacon frames following a random delay based on the contention window. Using the beacon signals, clocks synchronize with the fastest clock in the neighborhood to achieve time convergence. Such a method allows clocks to move forward but never backward essentially constituting a unidirectional clock~\cite{Zhou2007}. 
In both infrastructure and ad-hoc communication, the TSF-based synchronization technique sends messages by using the communication bandwidth of the network. This is known as in-band synchronization or internal synchronization~\cite{Tyrrell2010}.

\begin{figure}[tb!]
	\begin{center}
		\includegraphics[width=0.6\columnwidth]{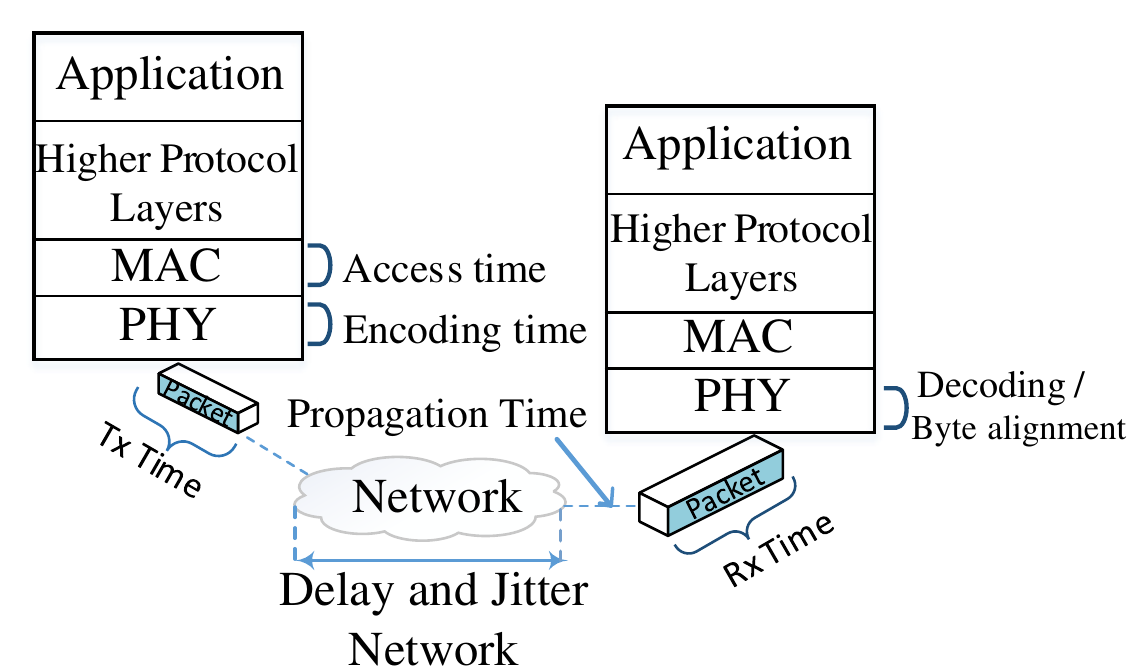}
	\end{center}
	\caption{Layered architecture of in-band time synchronization.}
	\label{Figure1}
\end{figure}

Such in-band synchronization not only consumes bandwidth that could be otherwise used for actual data transmission. It also causes message synchronization to experience significant jitters and delays, which will cause synchronization delays. \figurename~\ref{Figure1} shows the possible layer-level delays and the source of network jitters from in-band communication.
Such delay and jitters budget in communication networks have been extensively studied by Chen et al. \cite{chen2015implementation}, Abbas et al. \cite{Abbas2020}, Loschmidt et al. \cite{Loschmidt2012}, and Guenther et al. \cite{Guenther2005}. In vehicular networks, such end-to-end delays even increase with the number of nodes. According to Alonso and Mecklenbrauker, these delays can be up to 750 ms for a group of 50 nodes and 2s for a group of 200 nodes \cite{Alonso2012}. Similar results are also demonstrated by Bilstrup et al. \cite{Bilstrup2010} and Ebner et al \cite{Ebner2002a}. 
Therefore, time synchronization using this in-band TSF model is a challenge to achieving highly accurate time. It also takes longer to synchronize, posing an additional risk for safety message dissemination with a shorter life span. According to Cheng et al. \cite{cheng2006wireless}, the average maximum clock drift with TSF is 124.5 $\mu$s for 20 nodes. It raises to 500.2 $\mu$s when the number of nodes becomes 60. In addition, these 802.11p-based methods do not work beyond a single broadcast domain. This is also a serious limitation for their applications in vehicular networks.

Since TSF is solely responsible for the coordination of channel access of the wireless medium, it is still not intended to be used by the application layer of the system or for any other system and application purposes. In order to support time synchronization across  layers and applications, an amendment to IEEE 802.11 in 2012 proposes a method called Timing Advertisement (TA) to achieve synchronization.

\begin{figure}[tb!]
	\begin{center}
		\includegraphics[width=1.0\columnwidth]{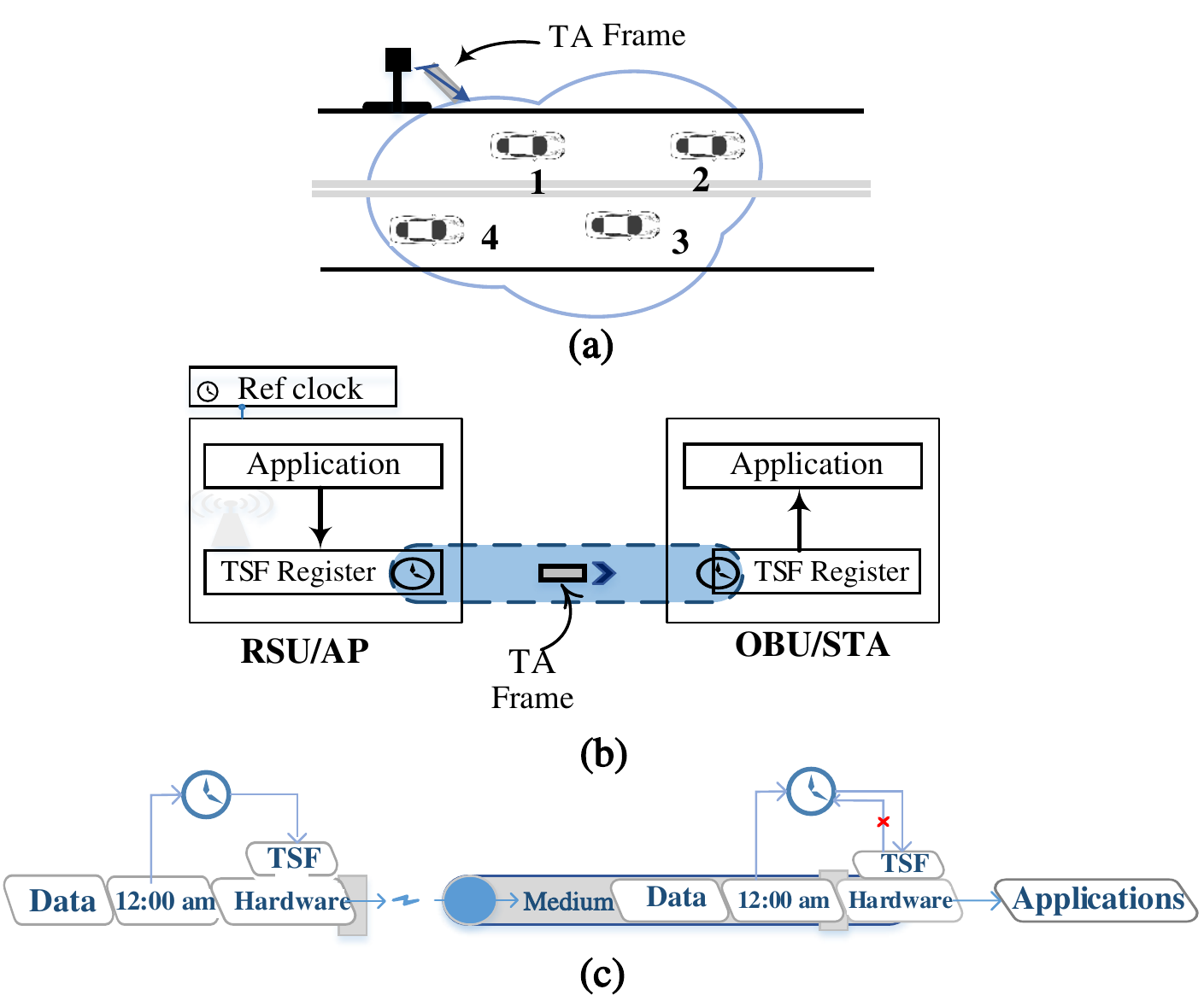}
	\end{center}
	\caption{Problem with TA-based time synchronization: (a) BSS Communication, Road Side Unit (RSU) sending beacons containing TA frames to synchronize; (b) A TA frame is transmitting from RSU to OBU; and (c) time development (transfer) in TA process.}
	\label{tymadvert}
\end{figure}

In the mechanism of Timing Advertisement (TA), the AP conveys a beacon that contains a time advertisement frame involving both the time of the local clock (i.e., 'time value') and the offset between the local clock and the global standard of time (i.e., ’time error’). The beacon or the probe messages can also use these two parameters to introduce time synchronization under a single BSS. An STA node revives the signal and updates the timing information sent across as shown in  \figurename{\ref{tymadvert}}. However, it is clear that such a synchronization mechanism depends entirely on the in-band data communication between the AP and STAs. Thus, it not only relies on the accurate computational performance but also creates overhead to the transferable payloads. In addition, in VANET a significant portion of network nodes are to be considered as ad-hoc V2V communication. This is compared with infrastructure-based V2I communication that resembles the AP to STA where the synchronization signal is steered by the infrastructure which makes VANET vary significantly from WLAN and difficult to achieve synchronization.
Therefore, although the infrastructure based V2I interaction has resemblance with WLAN communication paradigm where TA based time synchronization can be envisages as the mechanism to steer synchronization signal from the infrastructure to node, the ad-hoc portion (V2V) of VANET, however, vary significantly where it is difficult or not possible to achieve such infrastructure based synchronization signal.

Along with the time transferring primitives TA, the IEEE 802.11 is amended in 2012 with another mechanism known as the Timing Measurement (TM) frame. It allows individual STA (i.e. communicating nodes) to measure the relative offset between STAs. 
However, it is unclear how this relative time measurement is calculated or what degree of accuracy is possible to achieve.
In the amendment to IEEE 802.11 in 2016, however, a new frame named Fine Timing Measurement (FTM) is introduced. It aims to maintain the estimations of relative time offsets between AP and/or STAs with a certain high level of accuracy. It recommends that this time offset be achieved through Round Trip Time (RTT) calculation by regular transfer of message between communicating nodes. In view of communication modes, Timing Advertisement works on forward broadcasting messages whereas Fine Timing Management (FTM) is based on the back and forth message transmission mechanism among network elements. It is unclear under what circumstances which method will synchronize the network. In addition, it is still difficult to evaluate the performance of both TA and TM methods because no reports on this topic have been found in literature \cite{Mahmood2017}.

\section{Experimental Verification of GNSS Time Synchronization} \label{GNSStimesynch}

Generally, all GNSS systems maintain their timing accuracy and precision using expensive, accurate atomic clocks on-board in their satellite systems. they provides precise time benchmarks for applications through GNSS receivers. The everyday purpose GNSS receivers, however, are inexpensively made up with a less accurate inbuilt clock. Over time, yet, multiple techniques have been developed to transfer GNSS reference time to the earth-based receivers. Typical examples are One Way (OW), Common View (CV), All in View (AV), GNSS carrier Phase Time Transfer (CPTT), which are a few prominent standard timing techniques~\cite{Lewandowski1991,Dach2002,Petit2007,Defraigne2007b,Lewandowski1993}. One-way time transfer refers to the time dissemination where application receivers deduce GNSS time directly from GPS constellation, meaning that no additional techniques are not employed, e.g., comparison with another time from another receiver on earth, and same satellites (i.e., Common View (CV) and All in View (AV) technique).

In our experiments, we adapt the One Way time transfer technique. In most cases, the other time transfer techniques are not supportive or compatible with any systems that are in speed, such as high mobility based networks like VANET \cite{Melgaard1995}.  In the following sections, we will present the development details of our test-bed. Then, we will analyze our experimental results systematically under different road conditions.

\subsection{Development of our Experimental Test-bed } \label{experimentaltestbed}
The experimental test-bed is developed on a Single Board Computer (SBC) system, Raspberry Pi-3, Model B (Rpi-3B). SBC is convenient with its simplicity, programming capability and low power consumption characteristics ($<$5 W). It is small in size, and lightweight thus can be easily deployed for field operation. We did not use DSRC device as it is not programmable as we required for the test. Therefore, Raspberry Pi is considered as the alternative platform of DSRC based communication system for its simplicity, yet additional computing capabilities along with the availability of WiFi services onboard. Such low-level features offer relevance to the DSRC-based onboard communication system designed for vehicular communication. A part of the experimental demonstration uses WiFi as the transmission media to transmit the packet from one GNSS synchronized node to another. However, in such occasion, it is not used WiFi-based time synchronization but the proposed GNSS-based time synchronization approach. Since both WiFi (IEEE 802.11n) and DSRC (IEEE 802.11p) belong to the same wireless family (IEEE 802.11), to a large extent, their characteristics are similar. Therefore, in our experiment, any message transmission-related issue deploying WiFi should be similar to DSRC and is considered negligible in this scope of the experiment.

\begin{figure}[tb!]
\centering
\begin{subfigure}{0.40\columnwidth}
\includegraphics[width=1.0\linewidth,trim=0 15 170 5,clip]{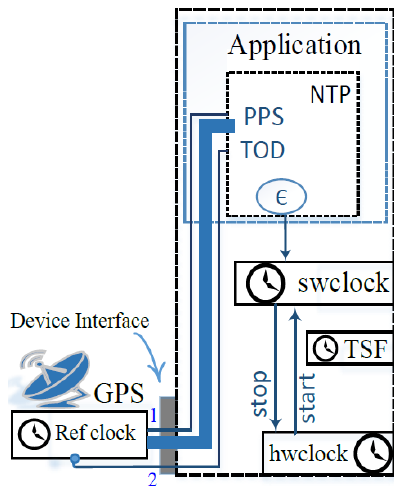}
\caption{Schematic diagram}
\end{subfigure}
\hfill
\begin{subfigure}{0.58\columnwidth}
\includegraphics[width=1.0\linewidth,trim=120 15 0 5,clip]{SchematicDiagramofDevelopment.pdf}
\caption{Laboratory set-up}
\end{subfigure}
	\caption{Test-bed for experiments.}
	\label{SchematicDiagramofDevelopment}
\end{figure}

RPI has some additional features. It provides a General Purpose Input Output (GPIO) port as a remarkably low-latency interfacing terminal, enabling direct integration of an external GPS module in our experiment. We interface a PPS-enabled GPS receiver from UBLOX (Model: Ublox MAX-M8Q) (\figurename{\ref{SchematicDiagramofDevelopment}}). Such a GPS receiver has two distinctive outputs: GPS data which contains NMEA sentences, and PPS pulse signal, as indicated by the interface terminal 1 and 2 in the \figurename{\ref{SchematicDiagramofDevelopment}}a. Since Rpi supports Linux-based operating systems, we power it with Jessie (Kernel version: 4.24), which is a lightweight operating system for low latency services.

\begin{figure}[tb!]
	\begin{center}
		\includegraphics[width=0.70\columnwidth]{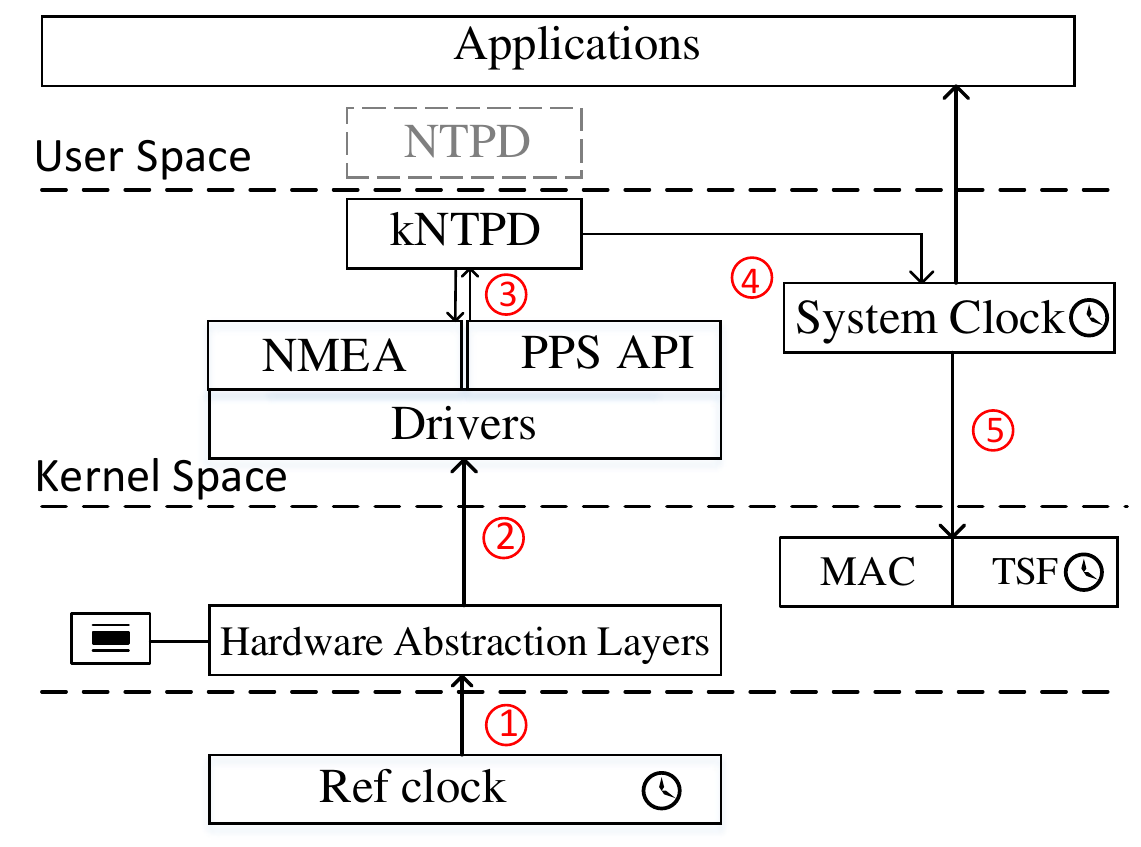}
	\end{center}
	\caption{A layered view of drivers and clocks in the kernel space of the operating system used in our experiments.}
	\label{LayerviewoftheDevelopment.pdf}
\end{figure}

To enable GPS-PPS in the system, a PPS API driver is installed. Its primary function is to connect the GPS receiver and timestamped external events deliver to a system with high resolution. We have patched the kernel of the operating system with the PPS-API patch to fetch the PPS signal from DCD pin of the serial port. The patch evaluates the offsets between the system clock and the PPS reference clock and passes it to the time daemon. Additionally, in this experiment, we have compiled the kernel of the operating system to fetch the drivers closer to the daemon as shown in a layer-view diagram \figurename{\ref{LayerviewoftheDevelopment.pdf}}.

To lock the system clock with a reference clock and to monitor the comparative results with other various timing statistics such as offsets and jitters, a time daemon is used in our experiments. There are a few time daemons that can be used for our experiments. We have employed the Network Time Protocol Daemon (NTPD) in our monitoring framework. NTPD is sensitive to nano-second scale measurements, and also to the availability of the connection. Therefore, when GPS signals are disrupted, NTPD shows deflection in output. 

In addition, we can record and store all statistical data by generating log files. The PPS pulse  does not contain any information on the absolute time (i.e. second, minute, hour, day, month, year) but just tick in every second to mark the beginning and/or ending of an arbitrary second. Therefore, we use a GPS NMEA driver to receive NMEA sentences to get real physical time. Then, we discipline the received time using PPS pulse. An alternative way is to use a wall clock from the Internet or from any other sources, but to ensure independent time in the node we used NMEA from the receiver. \figurename{\ref{LayerviewoftheDevelopment.pdf}} depicts the overall architecture of this experimental setting including how the reference clock is integrated and updates the system clock.


To evaluate and analyze the performance of GNSS time integration into our developed node, we have conducted a series of laboratory-based tests to characterize the timing performances of the GNSS receiver. We have also deployed the developed node for field experiments. Finally, we have synchronized three nodes with GNSS time to determine relative timing accuracy among them. The results and their analysis are presented below. 

\subsection{Performance of Node Clock Synchronization}
\begin{figure}[b!]
	\begin{center}
	
		\includegraphics[width=.80\columnwidth,trim=0 310 10 0,clip]{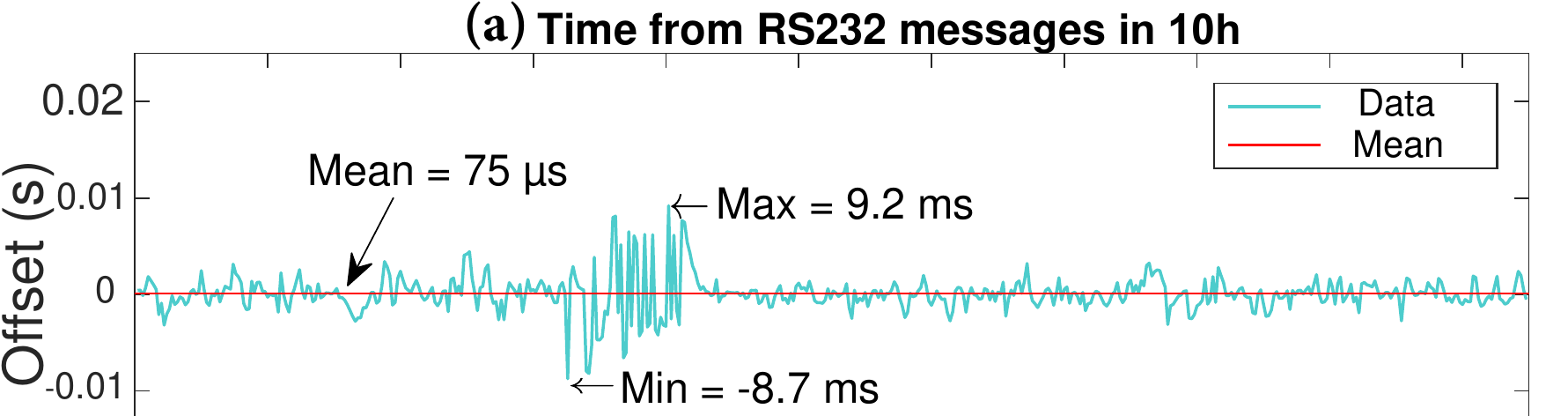}\\[4pt]
		\includegraphics[width=.80\columnwidth,trim=0 0 10 170,clip]{Figure-2.pdf}		
	\end{center}
	\caption{
		Timing synchronized with GPS signal: (a) a test run for 10 hours with temperature controlled at 16$^\circ$C; and (b) a text run for 24 hours with varying temperature between 20 - 25$^\circ$C.}
	\label{Figure3}
\end{figure}

In our laboratory test, from the message of the serial communication port RS 232 of the GPS receiver, the time tag is retrieved. It is also extracted separately from the PPS signal received from the available DCD pins of the same port. It is noticed that a number of commercial GPS receivers do not provide PPS signal output. Without PPS, it requires investigations whether or not the attainable timing accuracy from GPS NMEA data meets the application requirements. The timing information received from RS232 message through UART port is plotted in \figurename~\ref{Figure3}. It is seen from \figurename~\ref{Figure3}(a) that the recorded offsets of time are limited within 10 ms when the temperature is maintained constant at 16$^\circ$C.

The employed GPS receivers from U-Blox company is made up with Temperature Compensated Crystal Oscillator (TCXO), which has good stability over a broad range of temperatures (sub 0.1 ppm frequency stability over
an extended temperature range (50$^\circ$C to 105$^\circ$C)). It is, however, quartz-made clock where frequency variation is considered a commonplace due to the result of the environmental changes of the node, e.g., variations in temperature, pressure, and power voltage upon the clock \cite{sundararaman2005clock,Hasan2018}. Power voltage can be regulated to keep constant. The pressure is also constant in a certain territory. Thus, we have tried to record the effect of temperature changes on the timing performance of our developed node.

\figurename~\ref{Figure3} (b) shows 
timing information in our experiments of 24 hours in a normal room with temperature varying between $20^\circ - 25^\circ$C. The deviation of time due to temperature changes is obvious from the graph. The the standard deviation is recorded as 4.7$\mu$s in a normal room without room-temperature control. This is compared with a standard deviation of 1.88 $\mu$s in an air-conditioned room.
From the graph of \figurename~\ref{Figure3} (b), it is seen that the deviation of time is recorded as 5.2 ms over a period of 24 hours. 
From the data-set logged in room temperature, the maximum offsets recorded are +9.2 ms and -20.7 ms in positive and negative scales, respectively. 
The average value calculated as 5.2 ms is considerably high for many applications in network operation.

\begin{figure}[tb!]
	\begin{center}
		\includegraphics[width=0.80\columnwidth]{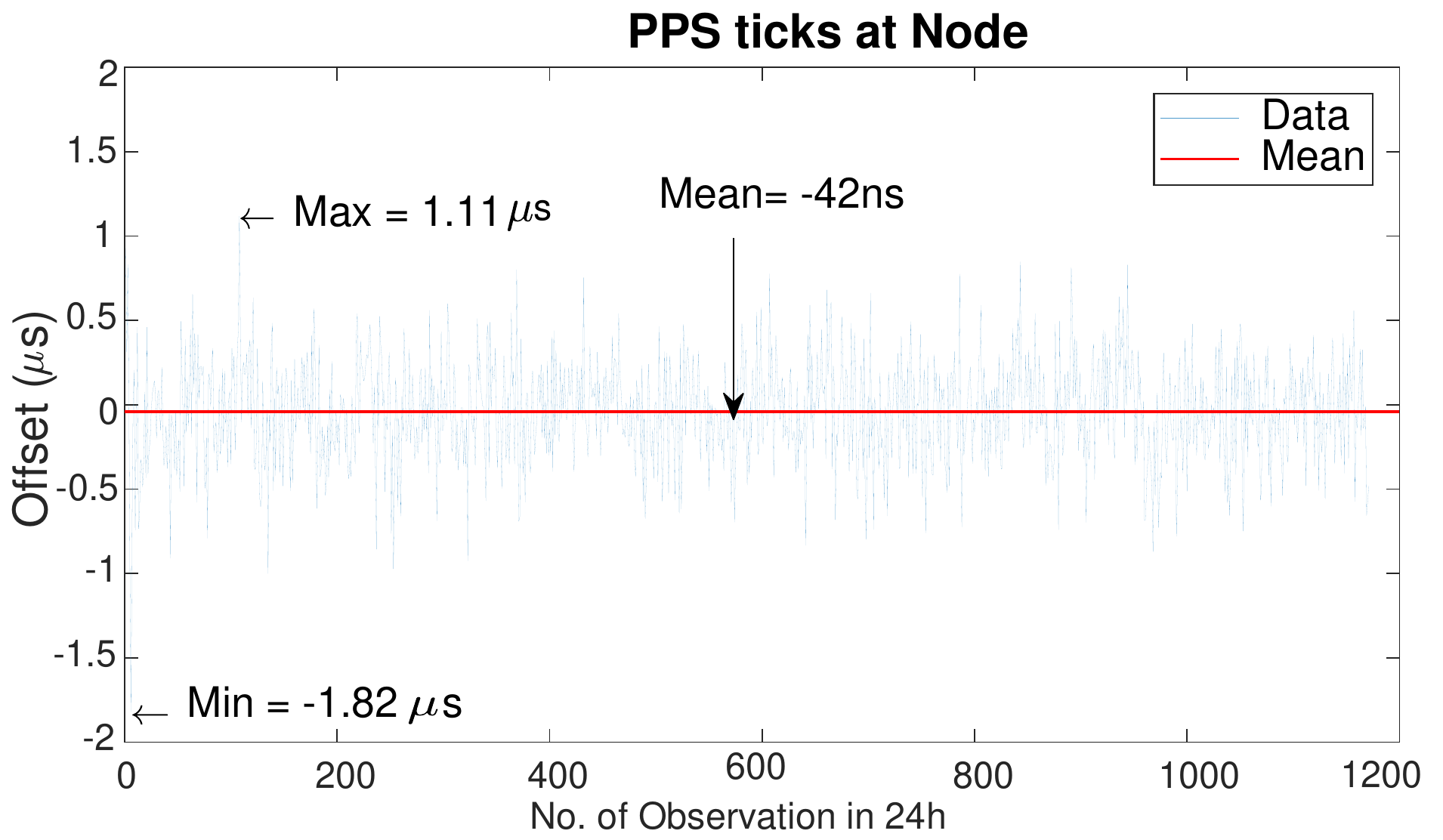}
	\end{center}
	\caption{
		System clock synchronized with PPS ticks.}
	\label{Figure4}
\end{figure}

\figurename~\ref{Figure4} presents the response of sole PPS signal. In compared with GPS data stream from RS232, it is seen that the PPS clock is highly accurate with an average offset of 42 ns. From a 24h observation, a maximum of 1.11 $\mu$s (1.82 $\mu$s in negative scale) is recorded. The vehicular network is a real-time network, where physical time plays a significant role in many applications.

\begin{figure}[tb!]
	\begin{center}
		\includegraphics[width=0.80\columnwidth]{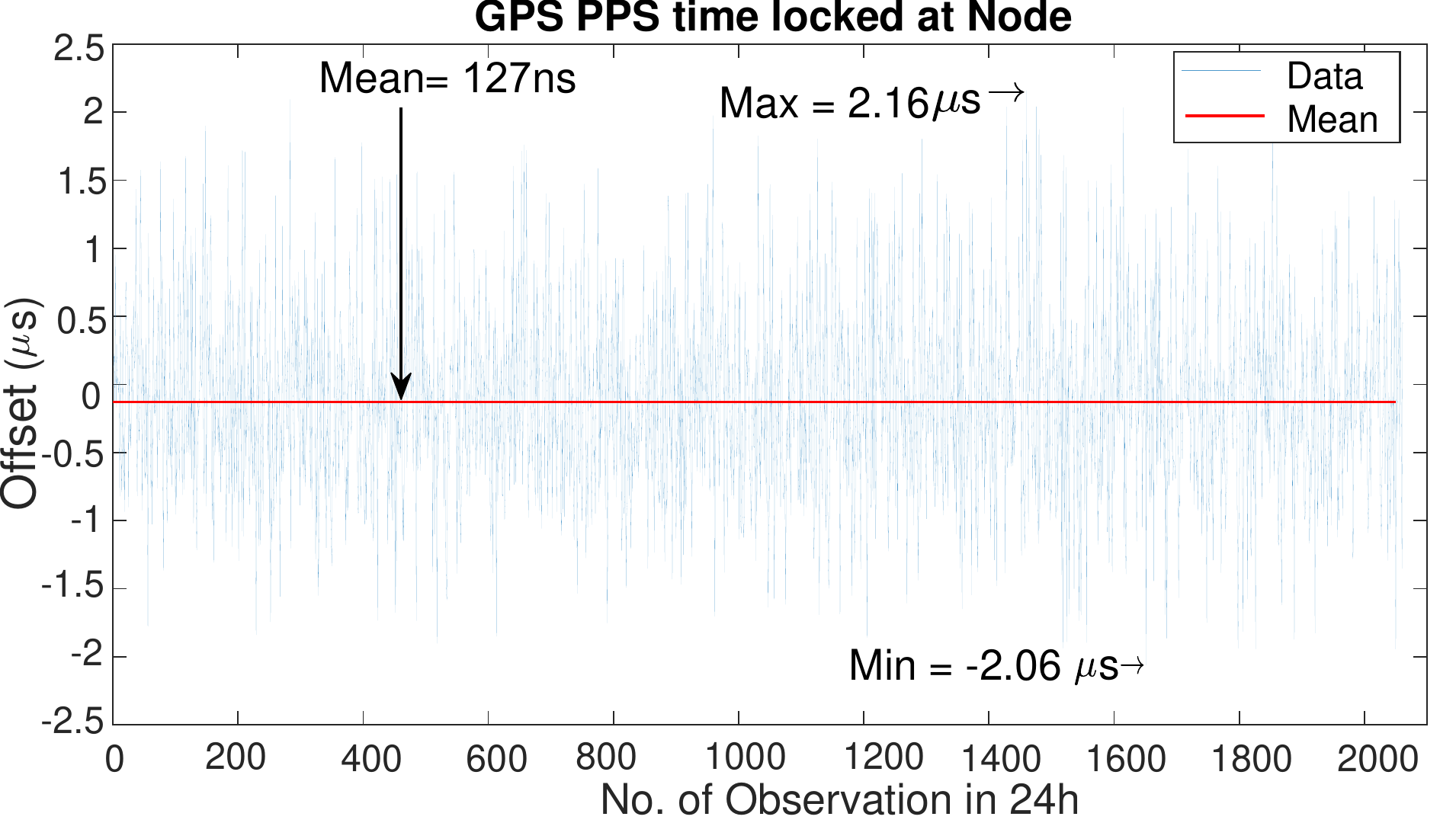}
	\end{center}
	\caption{\small 
		System clock synchronized with both GPS and PPS signals.}
	\label{Figure5}
\end{figure}

Although PPS tick is highly accurate, it does not carry any timing information.  
Therefore, a secondary source to signal is needed to get the physical time (i.e. time of day (TOD)). The GPS time tag supplied from the RS232 port can serve this purpose. 
Some researchers \cite{Guo2017,Ben-El-Kezadri2010,mangharam2007bounded} have conducted experiments to extract TOD timing information from the Internet, implying the requirement of network coverage with Internet facilities. In our experiment, however, we have used the TOD (i.e., UTC) received through RS232 port. Then, we combine this signal with with PPS signal. This arrangement makes system time dependent only on one system, i.e., GPS. No additional source of signals becomes necessary. 
\figurename~\ref{Figure5} shows the resultant clock performance of our GPS-PPS system over a period of 24 hours. It is seen from this graph that the Peak-to-Peak maximum offset is 4.22 $\mu$s (2.16$\mu$s+2.06$\mu$s). This means that two such networks can be synchronized with an accuracy of 4.22 $\mu$s maximum.

\begin{figure}[tb!]
	\begin{center}
		\includegraphics[width=0.8\columnwidth]{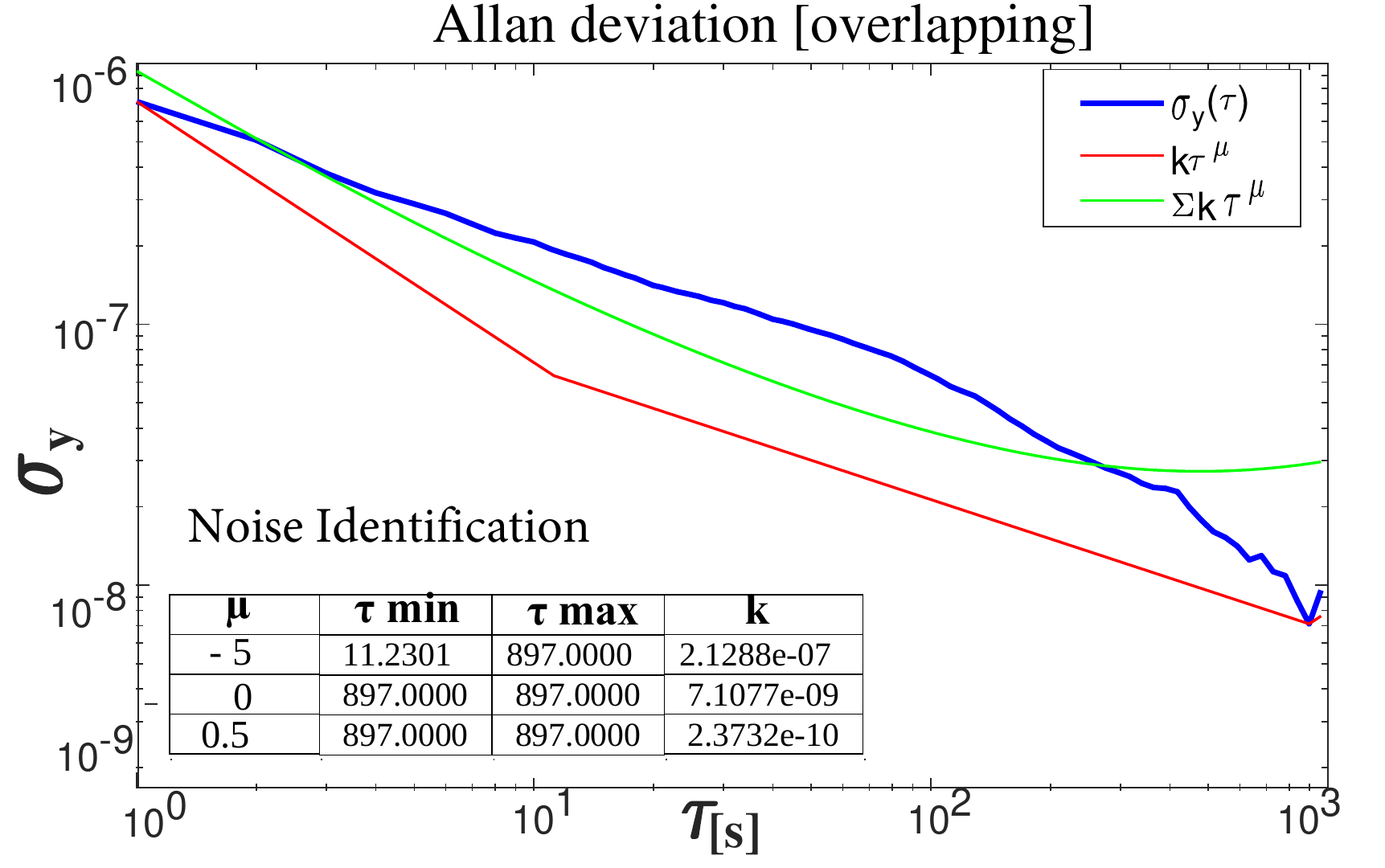}
	\end{center}
	\caption{Clock stability and noises using the Allan deviation.}
	\label{Figure7}
\end{figure}

To understand the clock stability and noise level, we have plotted the overlapping Allan Deviation ($\sigma$y) curve in \figurename~\ref{Figure7}. The noise template ($\kappa\tau^\mu$) slop and the sum of tangents ($\sum\kappa\tau^\mu$) are also plotted. From the slop analysis, White (thermal) noise ($\mu$=-.5), Pink (Electronics) noise ($\mu$=0) and Red (Angle rate) noise ($\mu$=.5) are all calculated.


\subsection{Field Test in Vehicular Environments}

\begin{figure}[tb!]
	\begin{center}
		\includegraphics[width=0.8\columnwidth]{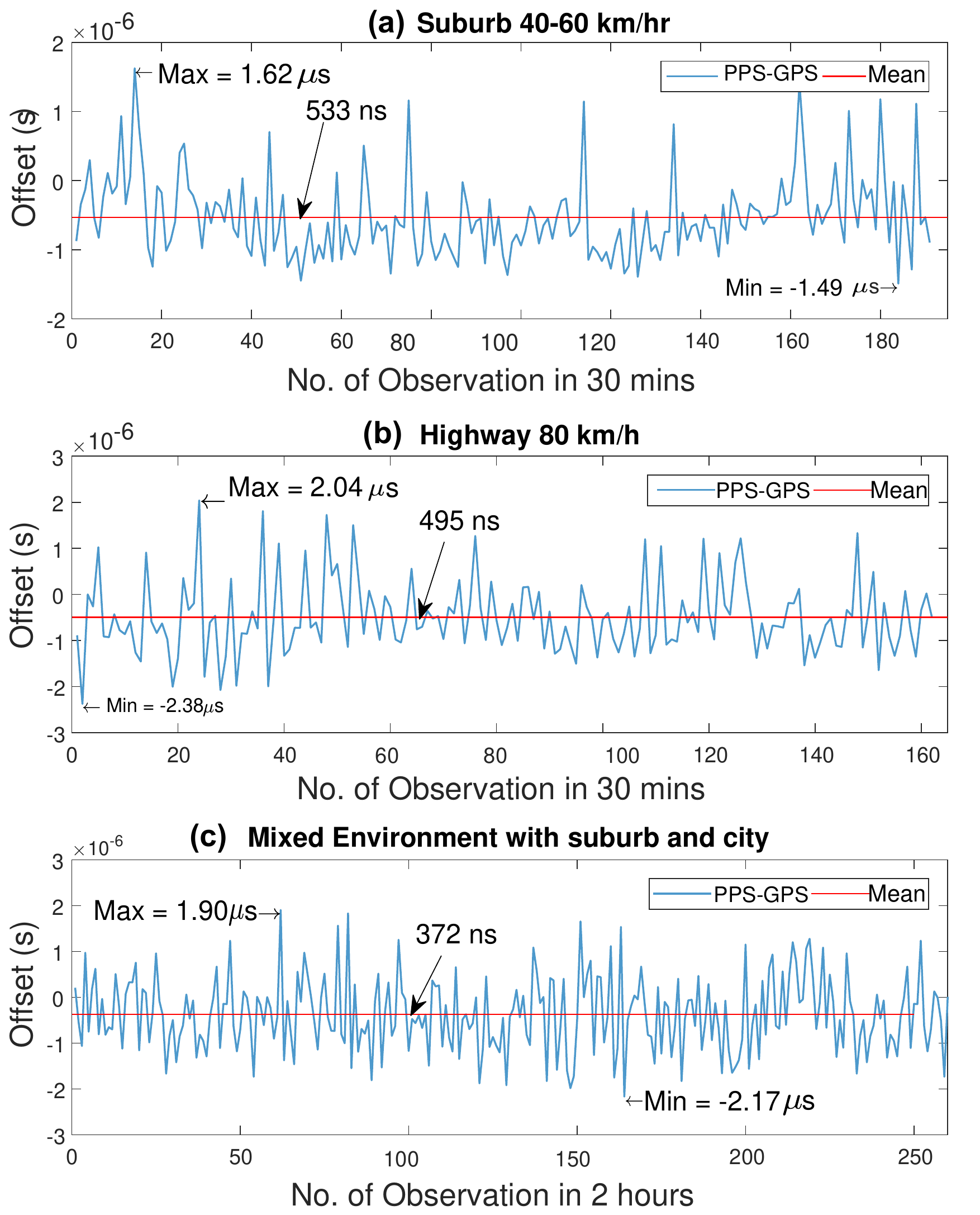}
	\end{center}
	\caption{\small 
	GPS-PPS-enabled clock in different road scenarios: (a) suburban roads 40 - 60 km/h; (b) highway 80 km/h; and (c) mixed roads in suburban and urban areas.}
	\label{Figure8}
\end{figure}

We have also conducted field experiments and recorded timing information in different road scenarios using our developed node. The results are plotted in \figurename~\ref{Figure8}. It is remarkable to notice that the developed clock performance on roads is even better than that from laboratory experiments. The maximum offsets between reference GPS and system clock are 1.62$\mu$s with an average of 533 ns over 30 minutes in suburban roads at a speed of 40 - 60 km/h. In highway (with a speed between 80-100 km/h), the maximum offset is 2.04$\mu$s with an average of 495 ns over 30 minutes. In a mixed suburban and urban roads with a varying speed between 40 to 60 km/h, the maximum of 2.17$\mu$s deflection is recorded with a peak-to-peak difference of 4.07$\mu$s maximum. We assume that due to the signal availability through RS232 data stream the overall performance is improved here. These results also indicate that the timing performance is not affected much by a moving speed of up to a 100 km/h.

\subsection{Network Time Synchronization}
This phase of the experiments investigates the timing accuracy between two GNSS-synchronized nodes, which are developed from our phase-one experiments discussed above. The experimental set-up is involved with one server as a monitoring node and two client nodes which are to be compared to get their timing accuracy. These two client nodes are connected over the symmetric link so that any data transmission related offsets from two GPS synchronized nodes to the third monitoring node become similar thus be cancelled out.

\begin{figure}[tb!]
	\begin{center}
		\includegraphics[width=0.5\columnwidth]{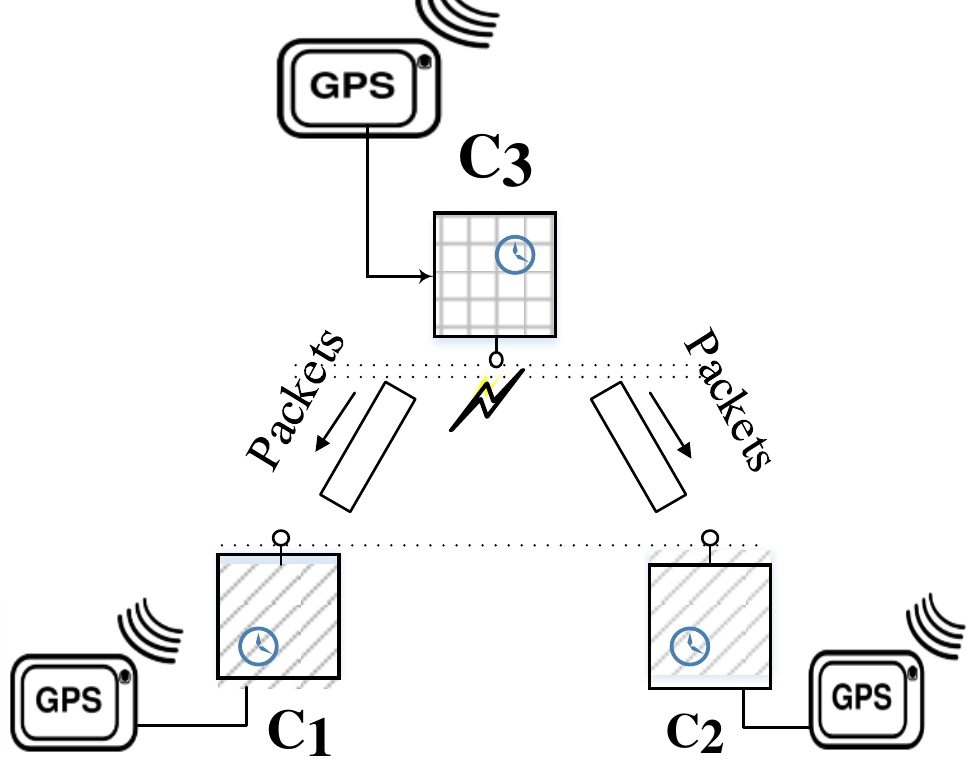}
	\end{center}
	\caption{\small 
	Experiment setup for measuring the time offsets between two GNSS-synchronized nodes.}
	\label{ExpsetupUdp}
\end{figure}

\figurename~\ref{ExpsetupUdp} depicts a server node $C_3$ sending UDP broadcast messages. These messages are received by clients $C_1$ and $C_2$, which are both synchronized with identical-model GNSS receivers. The central idea is that the server broadcasts messages to identically-configured receivers, which are assumed to receive at the same time. As they receive the message at the same time, the node’s individual time-stamps would yield their clock times.
In order to send and receive UDP messages, a simple \textit{socket program} is written in \textit{python}. The \textit{tcpdump} is used to record the time-stamps and compare the clock times. The experiment is implemented on the Raspberry Pi-3 with built-in wireless support (802.11n). Time-stamps are collected from the received packets of individual nodes in three datasets each with a different data rate: 10 packets per second, 100 packets per second, and 300 packets per minute.

\begin{figure}[tb!]
	\begin{center}
		\includegraphics[width=0.6\columnwidth]{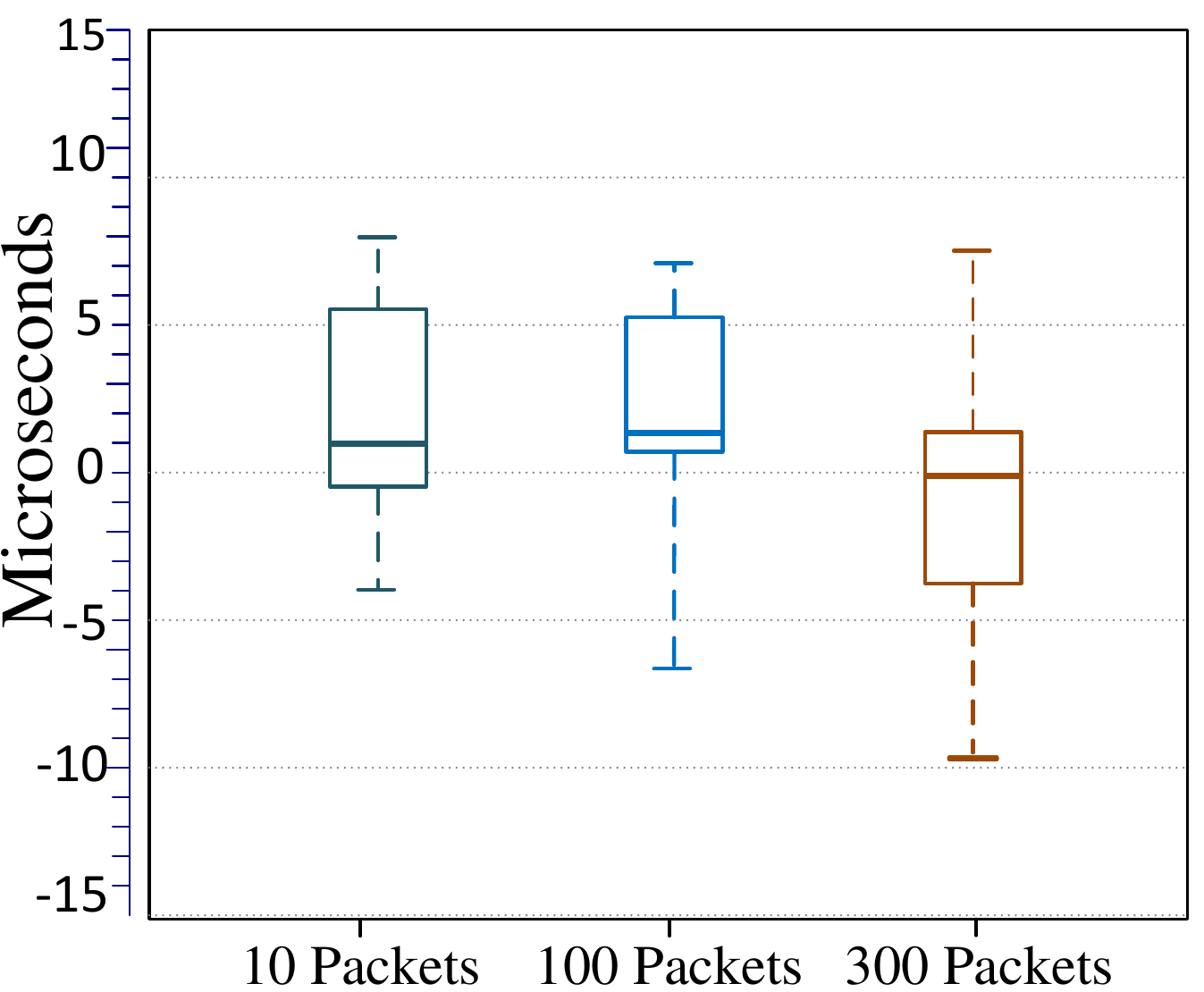}
	\end{center}
	\caption{\small Box-plots of time offsets between two GNSS synchronized nodes for three rates of data transmission.}
	\label{Figurebox}
\end{figure}

The results of our experiments are presented in box-plots  in \figurename~\ref{Figurebox}. For the 10-packet dataset, it is seen that there is an inter-quartile spread of approximately 6 $\mu$s, a median offset of 1 $\mu$s and a maximum offset recorded here is 8 $\mu$s. Similarly, the 100-packet dataset shows a 5 $\mu$s inter-quartile spread, a median offset of 1.5 $\mu$s and a maximum offset of 6.5 $\mu$s. The 300-packet dataset gives more variations with offsets ranging from 7.5 $\mu$s to 9.75 $\mu$s although 50\% of the (inter-quartile) values reside between 1.5 $\mu$s to -3.6 $\mu$s. The median value is minimal at 250 ns.

\subsection{Analysis on complete GNSS signal blockage}

With the standard positioning services (SPS) along with pseudo-range measurements, the accuracy and precision of GPS depend on the number of satellites visible at the ground-based receiver with a solid satellite geometry. Thus, the quality of service is vulnerable due to the signal unavailability in different obstructions on its line-of-sight (LOS) propagation ways under various challenging road scenarios such as large buildings, trees, tunnels, and bridges, etc. In such environments, the signal may be intermittent or entirely out of services for a while. 

For a better understanding of the challenging situation, a real-time vehicle-based experiment was reported previously in \cite{Hasan2018} over cluttered roads with a mix of high rising building, trees, short overpasses, and bridges in Brisbane city, Australia. 
This experiment developed an insight into the shielding of GNSS signals in different vehicular road environments by using consumer-grade GNSS receivers. With the advent of technologies, consumer-grade GNSS receivers were able to receive signals from multi-constellation GNSS systems. Out of four operational GNSS systems (i.e. GPS, GLONASS, BEIDOU, GALILEO), this experiment employed GPS and BDS (BEIDOU), implying an extended shielding service with full support of all these systems.  
The result reveals that the availability of both valid 4D positions and time solution for GPS+BDS is 49.6\%. Excitingly, the availability of 1 to 3 visible satellites are high, and a minimum 1 visible satellite (NSAT$\geq$1) in that area is 100\% in compared to 82\% with only GPS services. This means that with multi-GNSS service facilities, in any road circumstances there are almost always satellite signals.
Since the receiver can maintain time from the signal of one satellite, the availability of GNSS time solution is higher than the positioning solution in this challenging environment \cite{gps2006book,Hasan2018}. 

The experiment results presented in  \figurename{\ref{Figure8}} indicate and equivocal the findings of \cite{Hasan2018}. In the worst case, a positional error of 300 m will lead to 1 $\mu$s clock bias, the maximum deviation is recorded as 2.17 $\mu$s, meeting the timing requirements of vehicular networks~\cite{Hasan2018,FidaHasan2018,hasan2018gnss}.  

\begin{figure}[tb!]
	\begin{center}
		\includegraphics[width=0.5\columnwidth]{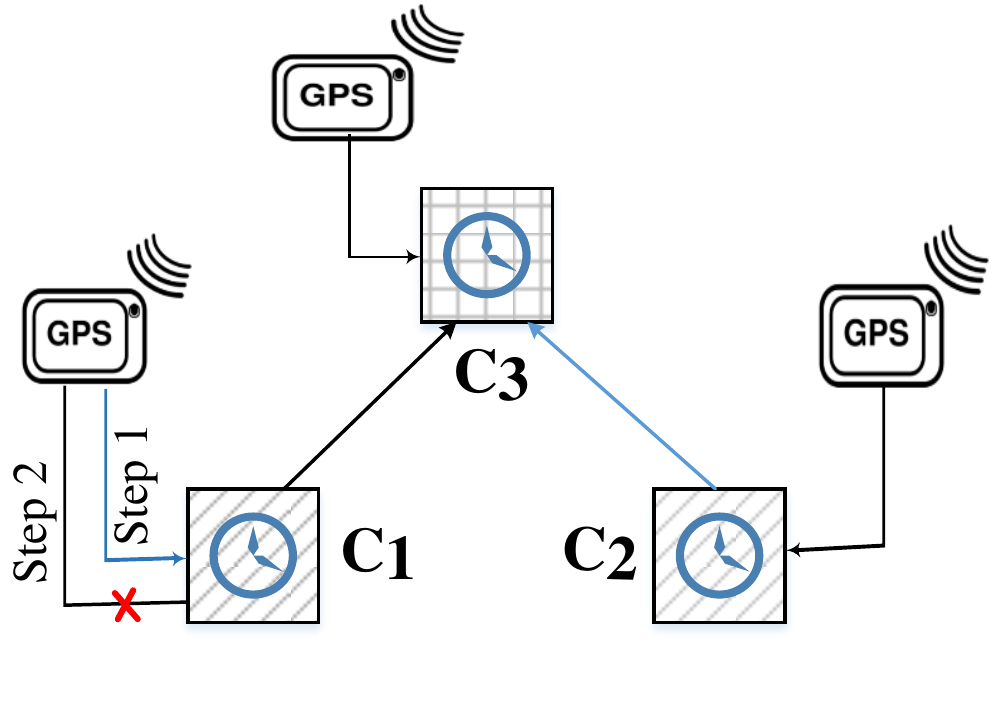}
		\caption{Experiment set-up with three nodes.
		}
		\label{Fig-6-6}
	\end{center}
\end{figure}

To understand the total blockage i.e., when NSAT=0, we have conducted another experiment.
In this experiment, three Raspberry Pi (Rpi-3) clocks are individually synchronized with three Ublox consumer-grade GNSS receivers as shown in \figurename{\ref{Fig-6-6}}. The node Clock $C_3$ is deployed as the server of the reference clock for comparing the time differences among the three GNSS synchronized nodes. 

\begin{figure}[tb!]
	\begin{center}
		\includegraphics[width=.8\columnwidth,trim=0 0 0 0,clip]{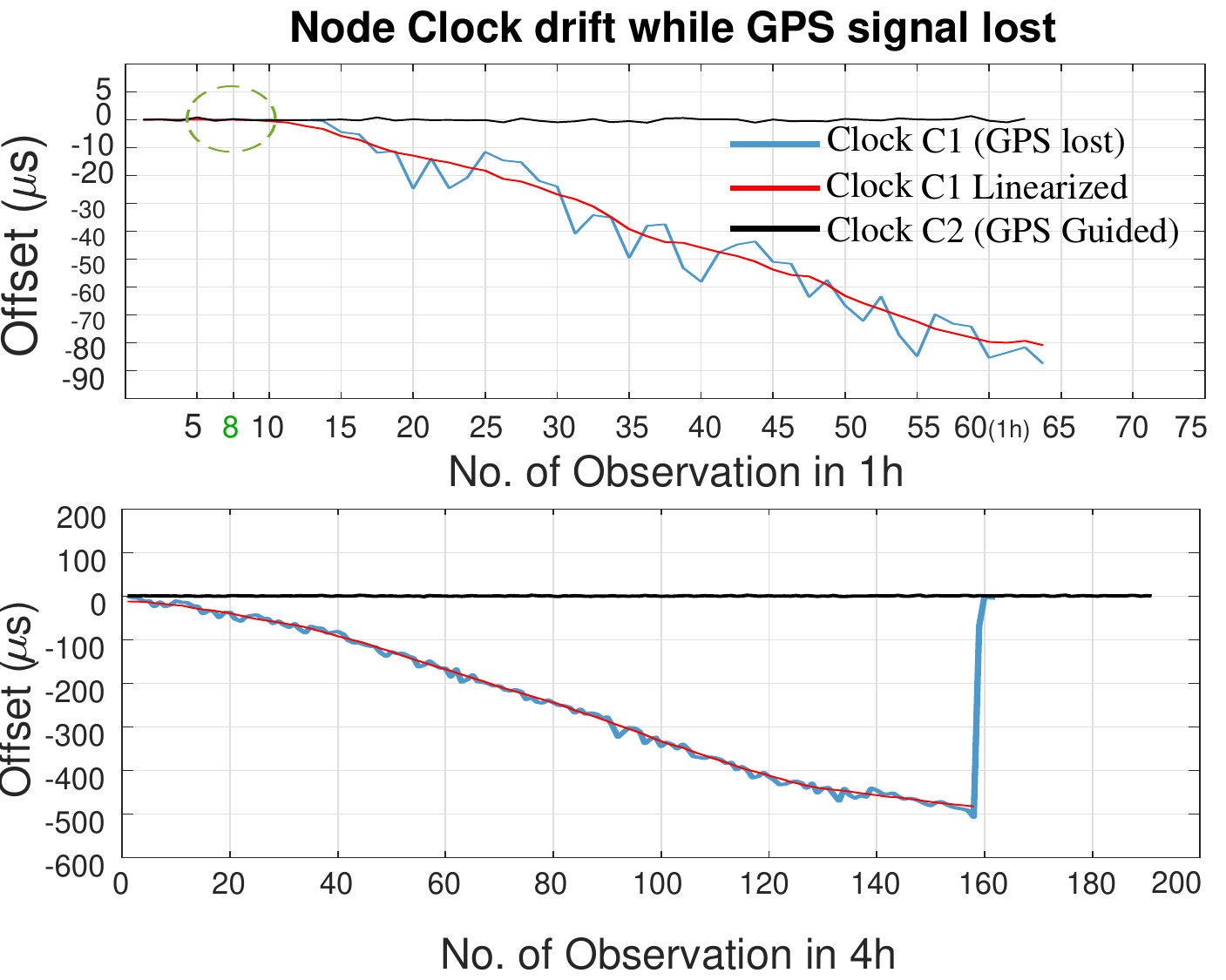}
		\caption{Clock drift recorded over 4 hours and 1 hour, reflectively.}
		\label{Fig-6-7}
	\end{center}
\end{figure}

At some point, the node Clock $C_1$ is disconnected from the GNSS antenna to mimic the signal loss. \figurename{\ref{Fig-6-7}} shows how the node clock $C_1$ is drifted without the presence of time-synchronization signal from GNSS. 
It is noted that at minute 8 the GNSS signal is disconnected and over the next hour, a drift of 80 $\mu$s is recorded. 
On a regular road condition, a signal loss for a maximum duration can occur under the tunnel. In Australia, the longest tunnel has a length of 5.25 km, for which a vehicle may stay out of GNSS services for about 5 min with a moving speed of 60 km/h. From \figurename{\ref{Fig-6-7}} (Top), a drift of 5 to 6 $\mu$s is observed, which still meets the vehicular network requirements.  

\figurename{\ref{Fig-6-7}} also shows a result of the clock drift over a 4-hour duration. The blue line represents the original drift with respect to elapsed time, and the red line is their moving average. From the moving average track, it follows that the trend of the drift is close to linear. This is an indication that the clock drift due to signal loss over a short period of time is approximately predictable. Thus, a predicted clock drift can bridge the timing outages when vehicles travel through tunnels where GNSS signals are completely blocked. 

In short, for some challenging road scenarios where there are fewer satellites in view, for example, when 1 to 3 visible satellites, GNSS developed clock can still give valid time solutions for accurate time synchronization in vehicular networks.
In the worst cases where the GNSS signals are entirely blocked, such as travelling in tunnels, the experiments validate that with such a complete outage for 5-6 minutes, the proposed solution could create a clock offsets of 7 to 8 microseconds. Therefore, GNSS approach is workable if the tunnels are within several kilometres in length, or the signal lost period is in the order of 5 minutes or shorter. For example, this solution is feasible in Australia, as the longest tunnel in Australia is 5.25 Km, which can introduce a signal lost period of fewer than 5 minutes. However, for an extended signal outage, the additional offsets can be further reduced using a predictive model, which is also achievable as demonstrated.


\section{Time Accuracy of LTE-based Cat M1 Service}

While DSRC (IEEE 802.11p) is considered the defacto connectivity solution to on-the-road V2X communication, LTE-based V2X is a competing connectivity solution~\cite{Araniti2013,Chen2016}. Recently, with the rise of the Internet of Things (IoT) technologies, Low Power Wide Area Networks (LPWAN) are getting much attention as another potential ITS connectivity solution. LPWAN includes two LTE-licensed cellular technologies, LTE CAT M1 (eMTC) and NB-IoT, which are able to support millions of devices per cell with a larger scale of coverage. Both technologies provide excellent security and quality of services (QoS). Compared with NB-IoT, CAT M1 offers extended bandwidth and provides a better data rate of up to 1 Mbps. It also supports mobility, suggesting that CAT M1 could be a viable ITS connectivity solution~\cite{Li2018}. 

There are already some studies focusing on DSRC-based synchronization as discussed previously. They are good references as they provide insights into the on-road timing capability of DSRC. The timing capability of LTE-based synchronization, however, has not been well understood. Therefore, we have conducted an experiment to synchronized a mobile node with an NTP server over the LTE-based cellular network. 

The idea behind this experiment is that the node periodically sends timing requests to an NTP server over LTE-based data communication service. The server also uses the same cellular communication service to sends its NTP responses. By using the sending and receiving timestamps, the traveling time (delay) and time differences (time offset) between the node and the server are estimated. Meanwhile, the node is synchronized with a precious GPS-enabled UTC supply. With a general algorithm, therefore,the UTC-enabled node calculates its time difference from the UTC time. 

\begin{figure}[tb!]
	\begin{center}
		\includegraphics[width=0.70\columnwidth]{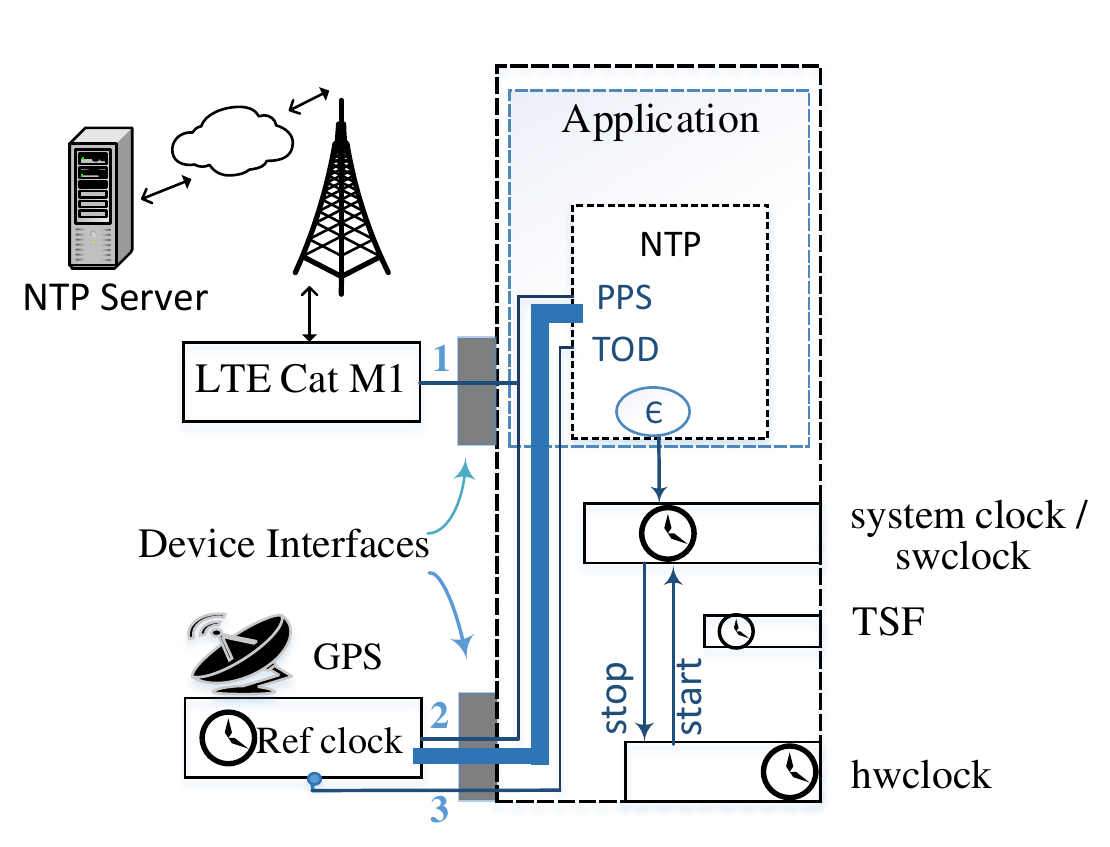}
	\end{center}
	\caption{\small 
		Test-bed to evaluate LTE-based time synchronization. 
	}
	\label{testbed}
\end{figure}

The testing node was developed on a single-board computer system, Raspberry Pi 3 (Model B). Raspbian ’Jessie (Kernel 4.14)’ and Linux PPS kernel module were installed on it. An LTE CAT M1 Raspberry Pi compatible HAT (Model: SIM7000E) was plugged to enable LTE-Based cellular network. The LTE service was provided by the cellular operator Telstra Corp., Australia. A PPS-enabled GPS Rpi compatible expansion board was also connected to the GPIO pin of the Raspberry Pi to get the precise UTC time as a reference for the timing comparison. A schematic diagram of the testbed is presented in \figurename{\ref{testbed}}. In the testbed, interface 1 represents LTE-based cellular connectivity using Cat M1 device to synchronize the node with an NTP server. With indicated interfaces 2 and 3, the node synchronizes with UTC time and PPS, respectively, to provide reference time for the system. The detailed setup including testbed development, necessary calibration, and the procedure employed for each of the testbed components has been described previously in Section \ref{experimentaltestbed}.
The overall set-up to the development is replicable as the hardware is commercially available, the parameter settings are given, and the obtained result from the setup does not rely on any unique laboratory or field-testing environment.

\begin{figure}[tb!]
	\begin{center}
	\includegraphics[width=0.75\columnwidth]{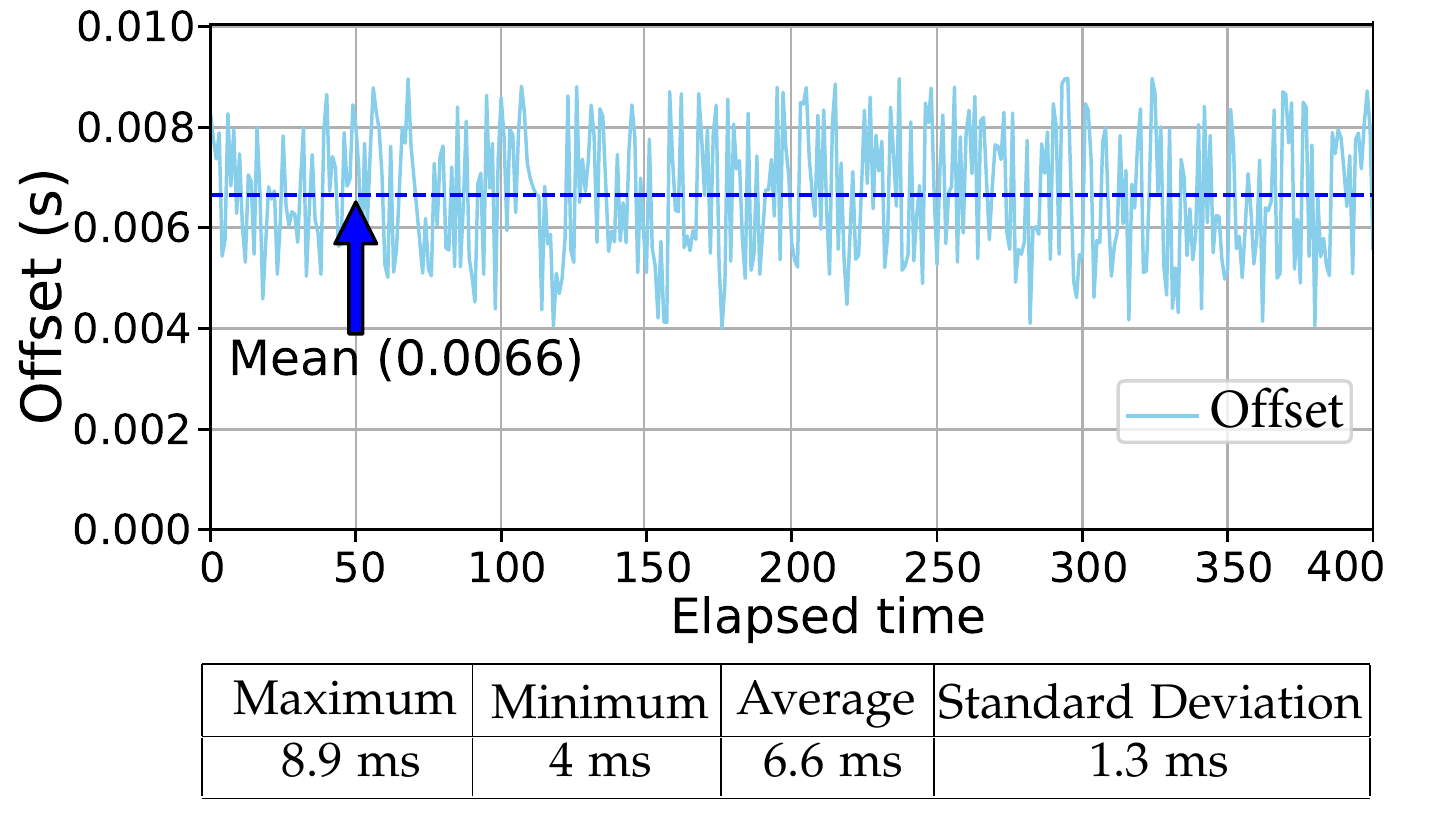}
	\end{center}
	\caption{Time offsets between GPS/UTC and NTP servers with an LTE link.}
	\label{LTEcatM1}
\end{figure}

In its operation, the developed mobile node sends NTP time synchronization requests via a cellular internet connection, receives responses, calculates offsets between the NTP time and its own time that has already been synchronized with the highly accurate GPS-PPS-supplied UTC time. The test results are plotted in \figurename{\ref{LTEcatM1}}. The recorded offset shows the maximum value of 0.0089 s to a minimum of 0.004 s with an average of .0066 s. This result implies that using LTE network time synchronization through an NTP server is as accurate as sub-10 ms. Since the NTP server is highly accurate with respect to UTC and is also steered by an atomic clock with a GPS receiver, the obtainable timing accuracy by a node via cellular network is not highly accurate. This is caused by the properties of cellular networks and also may be due to the asymmetry of the uplink and downlink communication protocols \cite{Miskinis2012, Miskinis2014}.  
Therefore, with the existing cellular network arrangement, the achievable time synchronization accuracy is not suitable for critical machine-type communication (cMTC), which is a targeted requirement for future cellular deployment \cite{Mahmood2019}.

\section{Conclusion}
This paper has presented a GNSS-based solution for stable and precise time synchronization in vehicular networks. From a systematic analysis of the time synchronization requirements for vehicular networks, this paper has outlined current synchronization practices with existing standard communication technologies such as DSRC and LTE-enabled vehicular networks. It also outlines their insufficiency while providing precise timing support to the highly-dynamic vehicular network.
Conducting a series of extensive experimental investigations, this paper demonstrated that the PPS-enabled GPS integration into an onboard vehicular unit (OBU) could achieve a tight bond with UTC at a peak clock deviation of 2.16 microseconds. In practice, such time synchronization accuracy in the vehicular networks cannot be easily achieved using other existing synchronization systems.


\bibliographystyle{IEEEtran}
\bibliography{bibliography_rev}

\begin{IEEEbiography}[{\includegraphics[width=1in,height=1.25in,clip,keepaspectratio]{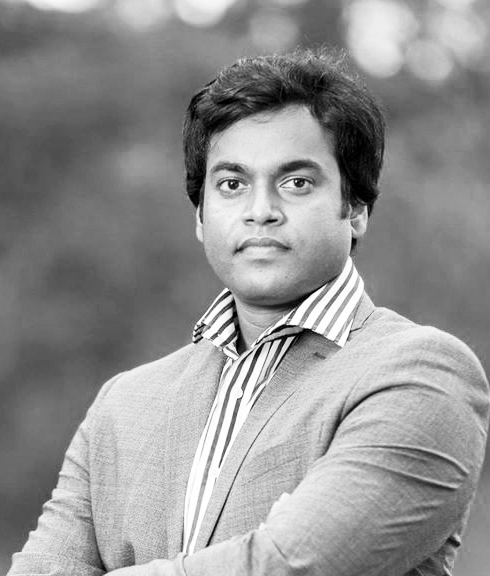}}]{Khondokar Fida Hasan} (S'13, M'18) Fida Hasan received a Ph.D. from the School of Electrical Engineering and Computer Science, Queensland University of Technology (QUT), Australia. He is currently working as a Postdoctoral Research Fellow at QUT. Fida Hasan awarded the Fellow of HEA, UK, for the excellence of his teaching and research practice in higher academia. His current research interests include Time Synchronization in the emerging Wireless Communications Networks including Intelligent Transportation Systems and Internet of Things. 
\end{IEEEbiography}
\begin{IEEEbiography}
	[{\includegraphics[width=1in,height=1.25in,clip,keepaspectratio]{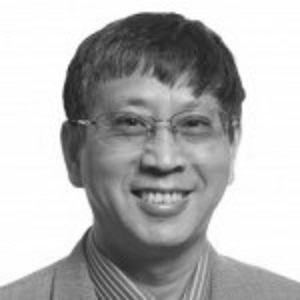}}]
	{Yanming Feng} received the Ph.D. degree in satellite geodesy from Wuhan Technical University of Surveying and Mapping (now part of Wuhan University), Wuhan, China. He is currently a professor at the School of Computer Science, Queensland University of Technology, Brisbane QLD, Australia. His research interests include satellite orbit dynamics, GNSS positioning, multiple-carrier ambiguity resolution, real-time kinematic positioning and applications, vehicular communications, and IoT. 
\end{IEEEbiography}
\begin{IEEEbiography}
	[{\includegraphics[width=1in,height=1.25in,clip,keepaspectratio]{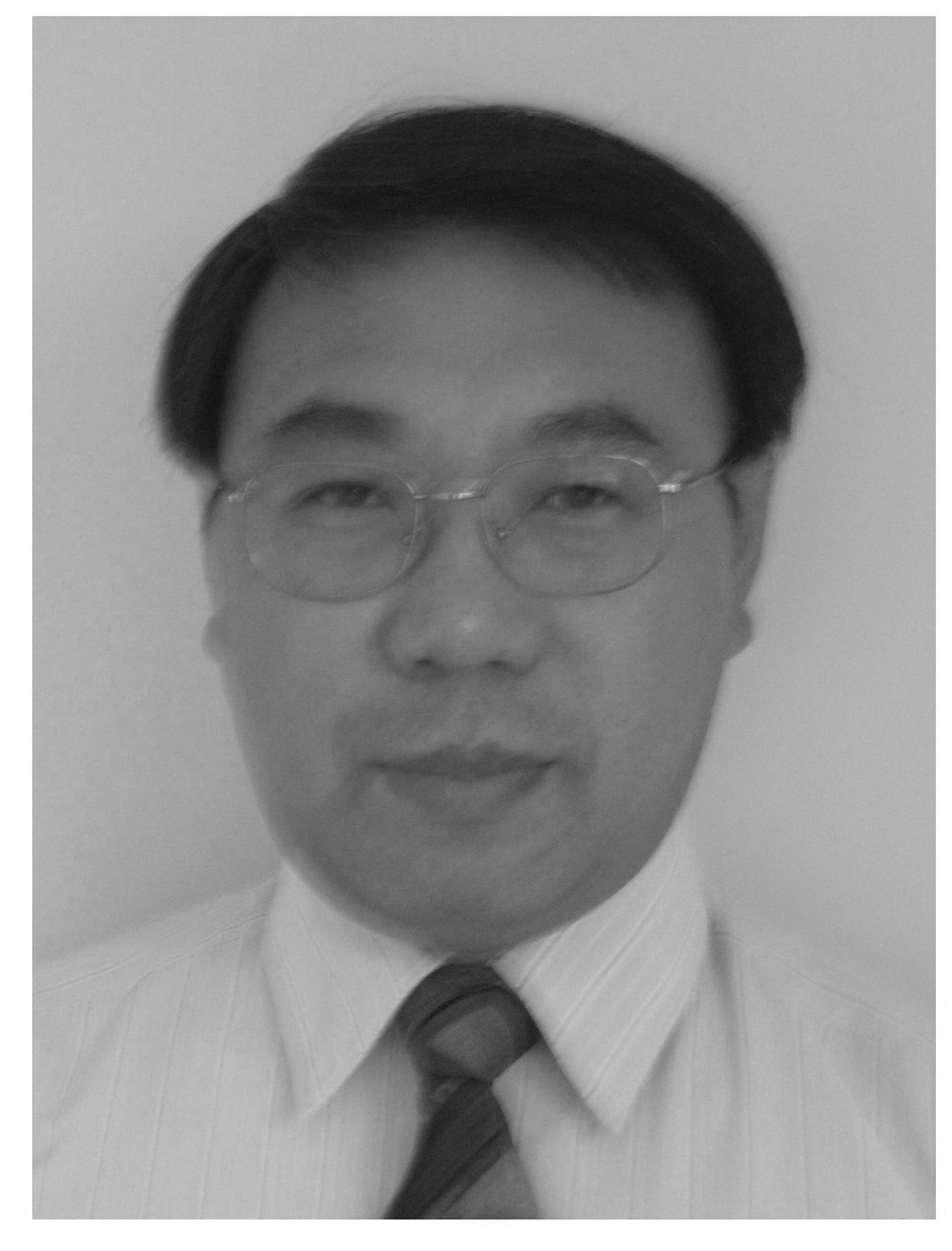}}]
	{Yu-Chu Tian} (SM'19, M'00) received the Ph.D. degree in computer and software engineering in 2009 from the University of Sydney, Australia, and the Ph.D. degree in industrial automation in 1993 from Zhejiang University, China. He is currently a professor at the School of Computer Science, Queensland University of Technology, Australia. His research interests include big data computing, cloud computing, computer networks, smart grid communications and control, optimization, artificial intelligence and machine learning, and control systems.
\end{IEEEbiography}
\end{document}